\documentclass[preprint,3p]{elsarticle}

\usepackage{graphicx}           
\graphicspath{ {Figures/} }
\usepackage{amssymb,amsmath}
\usepackage{xspace}
\usepackage{color}
\usepackage{longtable}
\usepackage{enumitem}
\usepackage{hyperref}
\usepackage{caption}
\usepackage{subcaption}
\usepackage{siunitx}
\usepackage{comment}
\usepackage{mathtools}
\usepackage[beginLComment=,endLComment=]{algpseudocodex}

\newcommand{\nc}{\newcommand}
\nc{\rnc}{\renewcommand}
\nc{\bs}{\boldsymbol}
\nc{\mrm}[1]{\mathrm{#1}}
\rnc{\matrix}[2]{\left[\!\!\begin{array}{#1}
	#2\end{array}\!\!\right]}
\rnc{\vector}[1]{\matrix{c}{#1}}
 \nc{\mm}[1]{\boldsymbol{#1}}
\nc{\real}[1]{\mathfrak{R}\left\{ #1 \right\}}
\nc{\imag}[1]{\mathfrak{I}\left\{ #1 \right\}}
\nc{\dd}{\mathrm{d}}
\nc{\ii}{\mathrm{j}}
\nc{\ee}{\mathrm{e}}
\nc{\inv}{^{-1}} 
\nc{\herm}{^{\mathrm H}}
\nc{\tra}{^{\mathrm T}}
\nc{\conj}[1]{ \overline{#1} }
\nc{\normal}{\mathrm n}
\nc{\tangential}{\mathrm t}
\nc{\kn}{{k_{\normal}}}
\nc{\kt}{{k_{\tangential}}}
\nc{\red}[1]{\textcolor{red}{#1}}
\nc{\green}[1]{\textcolor{green}{#1}}
\nc{\COMMENT}[1]{\textcolor{red}{#1}}
\nc{\MOD}[1]{\textcolor{blue}{#1}}
\nc{\ie}{i.\,e.\xspace}
\nc{\eg}{e.\,g.\xspace}
\nc{\vs}{vs.\xspace}
\nc{\cf}{cf.\,\xspace}
\nc{\myquote}[1]{`#1'}
\nc{\etal}{et al.\xspace}
\nc{\etc}{etc.\xspace}
\nc{\fabstand}{\,}
\nc{\fp}{\fabstand.}
\nc{\fk}{\fabstand,}
\nc{\msp}{\hspace{0.2cm}}
\nc{\fig}[4][tbh]{
\begin{figure}[#1]
\centering
\includegraphics[width=#4\textwidth]{Figures/#2}
\caption{#3\label{fig:#2}}
\end{figure}}
\nc{\e}[2]{\begin{equation} #1 \label {eq:#2} \end{equation}}
\nc{\est}[1]{\begin{equation*} #1 \end{equation*}}
\nc{\ea}[1]{
\begin{eqnarray}
#1 \end{eqnarray}}
\nc{\east}[1]{
\begin{eqnarray*}
#1 \end{eqnarray*}}
\nc{\fref}[1]{{Fig.~\ref{fig:#1}}}
\nc{\frefo}[1]{{\ref{fig:#1}}}
\nc{\frefs}[1]{{Figs.~\ref{fig:#1}}}
\nc{\freft}[2]{{Figs.~\ref{fig:#1}~and~\ref{fig:#2}}}
\nc{\tref}[1]{{Tab.~\ref{tab:#1}}}
\nc{\trefo}[1]{{\ref{tab:#1}}}
\nc{\trefs}[1]{{Tab.~\ref{tab:#1}}}
\nc{\eref}[1]{{Eq.~(\ref{eq:#1})}}
\nc{\erefo}[1]{(\ref{eq:#1})}
\nc{\erefs}[1]{{Eqs.~(\ref{eq:#1})}}
\nc{\ereft}[2]{{Eqs.~(\ref{eq:#1})~and~(\ref{eq:#2})}}
\nc{\erefm}[2]{{Eqs.~(\ref{eq:#1})-(\ref{eq:#2})}}
\nc{\sref}[1]{{Section~\ref{sec:#1}}}
\nc{\srefo}[1]{\ref{sec:#1}}
\nc{\srefs}[1]{{Sections~\ref{sec:#1}}}
\nc{\ssref}[1]{{Subsection~\ref{sec:#1}}}
\nc{\ssrefo}[1]{\ref{sec:#1}}
\nc{\ssrefs}[1]{{Subsections~\ref{sec:#1}}}
\nc{\aref}[1]{{{\ref{asec:#1}}}}
\nc{\arefo}[1]{{\ref{asec:#1}}}
\nc{\arefs}[1]{{{Appendices~\ref{asec:#1}}}}
\nc{\inst}[1]{$^{#1}$}
\nc{\naft}{N}
\nc{\ndof}{N_{\mathrm{DOF}}}
\nc{\Neq}{N_{\mathrm{eq}}}
\nc{\nfact}{N_{\mathrm{fact}}}
\nc{\nsolpt}{N_{\mathrm{pt}}}
\nc{\nord}{P}
\nc{\nnewtavg}{\overline{N}_{\mathrm{newt}}}
\nc{\regeps}{\varepsilon_{\mathrm{reg}}}
\nc{\epstol}{\varepsilon_{\mathrm{tol}}}
\nc{\sign}{\operatorname{sgn}}
\nc{\mex}{m_{\mrm{ex}}}
\nc{\dex}{d_{\mrm{ex}}}
\nc{\kex}{k_{\mrm{ex}}}
\nc{\omex}{\omega_{\mrm{ex}}}
\nc{\Dex}{D_{\mrm{ex}}}
\nc{\phex}{\bs{\phi}_{\mrm{ex}}}
\nc{\phdf}{\bs{\phi}_{\mrm{d}}}
\nc{\phnl}{\phi_{\mrm{nl}}}
\nc{\muex}{\mu_{\mrm{ex}}}
\nc{\eex}{{\bs e}_{\mrm{ex}}}
\nc{\edf}{{\bs e}_{\mrm{d}}}
\nc{\enl}{\boldsymbol{e}_\mrm{nl}}
\nc{\qex}{q_{\mrm{ex}}}
\nc{\dqex}{\dot{q}_{\mrm{ex}}}
\nc{\ddqex}{\ddot{q}_{\mrm{ex}}}
\nc{\kp}{k_{\mrm p}}
\nc{\ki}{k_{\mrm i}}
\nc{\zi}{z_{\mrm i}}
\nc{\dzi}{\dot{z}_{\mrm i}}
\nc{\kpnd}{\bar{k}_{\mrm p}}
\nc{\kind}{\bar{k}_{\mrm i}}
\nc{\err}{\epsilon}
\nc{\thref}{\frac\pi2}
\nc{\thf}{\vartheta_f}
\nc{\thfest}{\hat{\vartheta}_f}
\nc{\dthfest}{\dot{\hat{\vartheta}}_f}
\rnc{\th}{\vartheta}
\nc{\dth}{\dot{\vartheta}}
\nc{\thest}{\hat{\vartheta}}
\nc{\dthest}{\dot{\hat{\vartheta}}}
\nc{\thdel}{\vartheta_\Delta}
\nc{\thdelest}{\hat{\vartheta}_\Delta}
\nc{\Arg}[1]{\operatorname{Arg}\left\{ #1 \right\}}
\nc{\Uc}{U}
\nc{\delpl}{\delta_{\mrm{p}}}
\nc{\dellin}{\delta_{\mrm{s}}}
\nc{\delex}{\delta_\mrm{ex}}
\nc{\phexlin}{\phi_{\mrm{ex,lin}}}
\nc{\omlin}{\omega_{\mrm{lin}}}
\nc{\muexlin}{\mu_{\mrm{ex,lin}}}
\nc{\omLP}{\omega_{\mrm{LP}}}
\nc{\lamR}{\lambda_\mathrm{R}}
\nc{\lamI}{\lambda_\mathrm{I}}
\nc{\delplnd}{\bar{\delpl}}
\nc{\omnd}{\bar{\omega}}
\nc{\lamRnd}{{\lambda}_\mathrm{R}}
\nc{\lamInd}{{\lambda}_\mathrm{I}}
\nc{\errtol}{\err_{\mrm{tol}}}
\nc{\Omnd}{\bar{\Omega}}
\nc{\omexnd}{\bar{\omega}_\mrm{ex}}
\nc{\tnd}{\bar{t}}
\nc{\ff}{\bar{f}}
\nc{\ffd}{\bar{f}_\mrm{d}}
\nc{\uu}{\bar{u}}
\nc{\qq}{\bar{q}}
\nc{\FF}{\bar{F}}
\nc{\FFd}{\bar{F}_\mrm{d}}
\nc{\UU}{\bar{U}}
\nc{\QQ}{\bar{Q}}
\nc{\diag}[1]{\mrm{diag}\left\{ #1 \right\}}
\nc{\hs}[1]{\hspace*{#1}}
\nc{\x}[1]{\mbox{#1}}
\nc{\lk}{(\hs{1ex})}
\nc{\prog}[1]{{\sf{#1}}\xspace}
\nc{\name}[1]{\textsc{#1}\xspace}

\makeindex             
\journal{Elsevier}
\begin{document}

\begin{frontmatter}




\title{An iteration-free approach to excitation harmonization}

\author[UST]{Patrick Hippold}
\author[LUH]{Gleb Kleyman}
\author[UST]{Lukas Woiwode}
\author[ICL]{Tong Wei}
\author[UST]{Florian Mueller}
\author[ICL]{Christoph Schwingshackl}
\author[UST]{Maren Scheel}
\author[LUH]{Sebastian Tatzko}
\author[UST]{Malte Krack}

\affiliation[UST]{organization={University of Stuttgart},
            city={Stuttgart},
            country={Germany}}

\affiliation[ICL]{organization={Imperial College London},
            city={London},
            country={UK}}

\affiliation[LUH]{organization={Leibniz Universität Hannover},
            city={Hannover},
            country={Germany}}

\begin{abstract}
Sinusoidal excitation is particularly popular for testing structures in the nonlinear regime.
Due to the nonlinear behavior and the inevitable feedback of the structure on the exciter, higher harmonics in the applied excitation are generated.
This is undesired, because the acquired response may deviate substantially from that of the structure under purely sinusoidal excitation, in particular if one of the higher harmonics engages into resonance.
We present a new approach to suppress those higher excitation harmonics and thus the unwanted exciter-structure interaction:
Higher harmonics are added to the voltage input to the shaker whose Fourier coefficients are adjusted via feedback control until the excitation is purely sinusoidal.
The stability of this method is analyzed for a simplified model; the resulting closed-form expressions are useful, among others, to select an appropriate exciter configuration, including the drive point.
A practical procedure for the control design is suggested.
The proposed method is validated in virtual and real experiments of internally resonant structures, in the two common configurations of force excitation via a stinger and base excitation.
Excellent performance is achieved already when using the same control gains for all harmonics, throughout the tested range of amplitudes and frequencies, even in the strongly nonlinear regime.
Compared to the iterative state of the art, it is found that the proposed method is simpler to implement, enables faster testing and it is easy to achieve a lower harmonic distortion.
\end{abstract}




\end{frontmatter}


\section{Introduction\label{sec:intro}}
Vibration tests are carried out to identify, update and validate dynamic models of structures. 
While this is commonplace in the linear case, methods appropriate for the nonlinear regime are still under active research. 
Nonlinear behavior is caused, for instance, by frictional and unilateral interactions at contact interfaces, large displacements or rotations (geometric nonlinearity), and super-elastic or plastic material behavior.
Nonlinear behavior makes the response of a structure sensitive to the level of the applied excitation.
A well-defined excitation is therefore crucial for obtaining meaningful vibration measurements \cite{Ewins.2000,McConnell.2008}.
\\
As in the linear case, impact hammer testing is popular also in the nonlinear case \cite{Deaner.2015,Kuether.2016,Jin.2019}.
An important benefit is that the structure is free from any exciter influence during the ring-down.
However, because of the impulsive form of the applied excitation, many modes respond and generally interact in a nonlinear way.
This makes it difficult, if not impossible, to infer the behavior of individual modes from the measurements, nor to systematically investigate phenomena related to modal interactions.
For those reasons, impact hammer testing is limited to the rather weak nonlinear regime \cite{Allen.2010,Kuether.2016,Roettgen.2017}, and shaker-based excitation has to be used instead.
The shaker (or exciter) is attached via a stinger (also called push or drive rod), or the structure is mounted via a support structure on the armature or the slip table of a large shaker. 
In the former case, the force applied to the structure under test is measured and considered as the input (\emph{force excitation}).
In the latter case, the imposed base motion is measured and considered as input (\emph{base excitation}).
Force excitation can also be implemented in a non-contact way by using some kind of voice-coil actuator \cite{Denis.2018,Givois.2020,Schwarz.2023}.
\\
A sinus wave is the by far most popular command signal in nonlinear vibration testing, and typically the focus is placed on the frequency range around a particular primary or secondary resonance, where the nonlinear behavior is most prominent. 
Feedback control is employed to achieve and maintain a given response or excitation level, and the excitation frequency is either imposed or adjusted by another control loop to achieve a given phase lag between excitation and response \cite{Raze.2024}.
The testing procedures mainly differ by what quantity is fixed and what is stepped:
In a classical stepped-sine test, the excitation level is maintained while the frequency is stepped \cite{Ewins.2000}. 
This way, one obtains a \emph{frequency response curve}.
In the nonlinear case, this curve may feature turning points giving rise to multiple coexisting vibration states for the same excitation frequency and level.
To achieve robustness near the resonance peak and to test also the overhanging branch between the turning points (which is unstable in open-loop conditions), phase control is used \cite{Sokolov.2001}.
In a phase-resonance test, a resonant phase lag is maintained while the excitation or response level is stepped \cite{Ewins.2000}.
In a response-controlled test, the response level is maintained while the frequency is stepped, see \eg \cite{Karaagacl.2021}.
It is also common to fix the frequency and step the response level to obtain so-called S-curves, see \eg \cite{Abeloos.2022}.
\\
While it is state of the art to control amplitude and phase of the sinusoidal excitation, the distortion of the excitation waveform is only rarely addressed.
Any form of nonlinear behavior present in the coupled exciter-structure system generates higher harmonics.
The nonlinear behavior of the structure is of key interest in the present work, and is in this sense inevitable.
But also the exciter can be an important source of nonlinear behavior.
An example are large displacements of the coil within the non-homogeneous magnetic field of an electro-dynamic exciter \cite{Tomlinson.1979}.
When higher harmonics are generated only within the structure, higher harmonics will also be present in the applied excitation due to the undesired feedback of the structure on the attached exciter.
When higher harmonics are generated only within the exciter, they will of course also be present in the applied excitation.
In general, the higher harmonic content of the applied excitation is the result of the interplay between (possibly nonlinear) exciter and nonlinear structure.
In this sense, the response measurements are not simply the cause of a well-defined input, but they are contaminated by the behavior of the exciter.
This is an important impediment for the task pursued with the vibration test, namely to identify/update/validate a model of the structure, as opposed to a model of the exciter-structure system.
The detrimental effect of the higher excitation harmonics has been recognized in experimental studies, see \eg \cite{Claeys.2014,Chen.2016,Shaw.2016,Renson.2019,Pacini.2022}.
The situation is particularly critical when the structure features harmonically coupled modes, \ie, there is an internal resonance, or when secondary (external) resonances are analyzed \cite{Renson.2016,Zhou.2024} .
If the purpose of the test is to validate a model, one might be tempted to tolerate the harmonic distortion, and simply impose the measured multi-harmonic excitation on the model of the structure.
However, this approach ignores that the higher harmonics actually result from the exciter-structure interaction. 
As shown in \cite{Krack.2021}, a small model error may in this case lead to unbounded response errors \cite{Krack.2021}.
\\
Model-based and model-free techniques have been proposed to directly control the dynamic excitation signal \cite{McConnell.2008}. 
It is the present state of knowledge that those techniques are limited to the regime outside resonance, whereas the focus of the method proposed in the present work is precisely the resonance regime.
To robustly handle the resonance regime, a less dynamic strategy is pursued by focusing on periodic vibration states and operating on the Fourier coefficients instead of the dynamic (time-domain) excitation signal.
To cancel the unwanted higher harmonics in the applied excitation\footnote{The \emph{applied excitation} is the force applied to the drive point in the case of force excitation, and the base motion in the case of base excitation.}, higher harmonics are added to the command signal, \ie, the voltage signal fed to the (power amplifier of the) exciter.
This idea dates back at least to the late 1990s \cite{Bucher.1998,Josefsson.2006}.
All existing methods achieve the harmonization by the \emph{iterative solution of a root finding problem}, see \eg \cite{Josefsson.2006,Shaw.2016,Renson.2019,Tatzko.2023}.
More specifically, the entries of the residual vector are the Fourier coefficients of the excitation, and the entries of the vector of unknowns are the sought Fourier coefficients of the command signal.
The equation system is solved using a Newton-type iteration method.
Here, the exciter-structure system is treated as a black box, and the derivative of the residual with respect to the unknowns is approximated using finite differences.
To obtain frequency response curves, this approach can be embedded within an experimental path continuation technique (Control-Based Continuation \cite{Sieber.2008,Sieber.2008b}).
A critical aspect is the selection of an appropriate step size for the finite differences:
If it is too small, noise leads to poor accuracy.
If it is too large, non-linearity leads to poor accuracy.
In practice, this limits the achievable quality of the harmonization for a reasonable number of iterations \cite{Tatzko.2023}.
The iterative character of the method leads to a long test duration, increasing the risk of testing-induced (\eg fatigue or fretting) damage, and impeding the analysis of time-variable systems.
In fact, time-variability is an important aspect in the present work, where the doubly-clamped beam considered in the experiment showed high sensitivity to slight temperature changes, which arose over a long test duration.
\\
In the present work, an iteration-free method is proposed for the purpose of suppressing higher harmonics in the excitation.
The Fourier coefficients of the command signal are adjusted by a feedback loop, involving a steady-flow estimation of the Fourier coefficients of the excitation, and proportional-integral controllers.
The proposed method is explained in \sref{methodology}.
In \sref{num_vali}, the method is validated in a virtual experiment.
In \sref{exp_vali}, the robustness of the method is evaluated in a real experiment, and the method is assessed against the iterative state of the art.
Conclusions are drawn in \sref{conclusions}.

\section{Proposed approach\label{sec:methodology}}
\begin{figure}[h!]
    \centering
    \includegraphics[width=\textwidth]{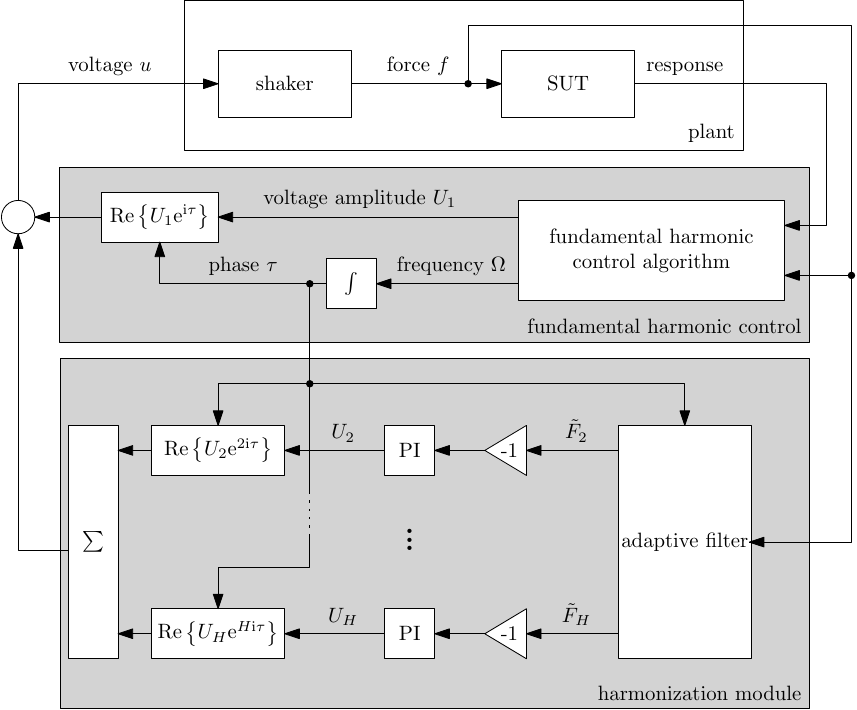}
    \caption{Conventional shaker-based sinusoidal vibration test (force excitation) extended by proposed \emph{harmonization module}. SUT: structure under test.}
    \label{fig:harmonization}
\end{figure}
%
An overview of the conventional sinusoidal testing and its extension by the proposed \emph{harmonization module} is given in \fref{harmonization} for the case of force excitation.
The command signal $u$ is the voltage fed to the (power amplifier of the) shaker,
\begin{equation}
    u = \real{U_1\ee^{\ii\tau}} +  \real{\sum \limits_{h =2}^{H} U_h\ee^{\ii h \tau}}\fk \label{eq:voltage}
\end{equation}
with the complex Fourier coefficients $U_h\in\mathbb C$ and the imaginary unit $\ii=\sqrt{-1}$.
$H$ is the truncation order of the harmonization.\footnote{For ease of notation, all higher harmonics up to $h=H$ are suppressed throughout this work. In practice, it may be useful to suppress only a subset, \eg, only the odd harmonics.}
The proposed harmonization should, in principle, be compatible with any state-of-the-art method for \emph{fundamental harmonic control}.
The fundamental harmonic control determines $U_1$ and $\Omega$.
The phase $\tau$ is the integral of $\Omega$, or equivalently, it holds that $\dot\tau = \Omega$, where over-dot denotes derivative with respect to time $t$.
\\
A proportional-integral controller is applied to each higher harmonic,
\ea{
\err_h &=& -\tilde{F}_h\fk \label{eq:errh} \\
U_h &=& k_{\mathrm{p}} \dot{I}_h + k_{\mrm i} I_h\fk \label{eq:Uh} \\
\dot{I}_h &=& \err_h \fp \label{eq:Ihdot}
}
Herein, $\tilde{F}_h$ is the $h$-th Fourier coefficient of the applied force $f$, estimated as described below.
Since the goal is to suppress higher harmonics, the set value is 0, so that the control error is $\err_h = 0-\tilde{F}_h = -\tilde{F}_h$.
$I_h$ is an auxiliary quantity (integral of control error).
Throughout the virtual and real experiments carried out in the present work, the same proportional and integral gains, $k_{\mrm p}$ and $k_{\mrm i}$, were used for each harmonic.
In some cases, an individual choice of $k_{\mrm p}$ and $k_{\mrm i}$ for certain harmonics can make sense.
\\
A steady-flow estimate of the Fourier coefficient $\tilde{F}_h$ is obtained as
\begin{equation}
    \dot{\tilde{F}}_h = 2\omLP \ee^{-\ii h\tau} \left(f-\real{\sum\limits_{h=0}^{H} \ee^{\ii h\tau}\tilde{F}_h}\right)\fp 
    \label{eq:AF}
\end{equation}
This is a time-continuous variant of Widrow's least-mean-squares (LMS) algorithm \cite{Widrow.1975}.
The LMS algorithm is commonly used to remove periodic disturbances from a given measurement (adaptive notch filter).
In this work, the goal is not to filter the measurement data, but to remove higher harmonic disturbances from the excitation applied to the structure under test.
Still, we refer to the procedure to estimate Fourier coefficients using \eref{AF} as \emph{adaptive filtering}.
Adaptive filtering was introduced to nonlinear vibration testing by Abeloos \etal \cite{Abeloos.2021}, and is known for its superiority over alternatives such as synchronous demodulation \cite{Abeloos.24November2022,Hippold.2024}.
The particular time-continuous form of the filter in \eref{AF} was proposed in \cite{Hippold.2024}.
The parameters of this adaptive filter are the order $H$ and the cutoff frequency\footnote{As shown in \cite{Hippold.2024}, in period-average, the adaptive filter acts as first-order low-pass with the cutoff frequency $\omLP$.} $\omLP$.
We recommend to set the order $H$ equal to the truncation order of the harmonization throughout this work.
For completeness, it should be remarked that the adaptive filter order must be greater or equal to the truncation order of the harmonization.
An estimate of the fundamental Fourier coefficient, $\tilde{F}_1$, is needed for the fundamental harmonic control.
Therefore, a single adaptive filter will be used in practice, for both the fundamental harmonic control and the harmonization module, whereas the adaptive filter is indicated as exclusive part of the harmonization module for the simplified overview in \fref{harmonization}.
\\
It is important to note that the proposed harmonization applies a controller \emph{individually to each harmonic}.
The underlying working hypothesis is that such a harmonically decoupled control is sufficient.
This working hypothesis will be carefully analyzed using virtual and real experiments.
\\
The key parameters of the proposed harmonization module are the gains $k_{\mrm p}$ and $k_{\mrm i}$.
A heuristic method to select those gains is proposed, as opposed to a model-based method, for the reason explained below.
Before the nonlinear test, we may assume that we have linear modal data, which would allow us to set up a linear model of the structure.
This is normally a good point of departure for model-based control design, see \eg \cite{Hippold.2024} for the case of fundamental harmonic phase control.
But this is impossible for the task of harmonization, because the generation of higher harmonics as well as the interaction among different harmonics are essentially nonlinear phenomena, \ie, it cannot be explained by linear theory.
As a consequence, a nonlinear model would be needed for model-based control design.
However, the very purpose of the test is to explore the nonlinear behavior; \ie, if we had a valid nonlinear model, there would be no reason to do the test.
Thus, the design of a higher harmonic controller based on a nonlinear model has no practical use.
\\
The remainder of this section is organized as follows.
First, we discuss a few fundamental relations between the parameters of controller, exciter and structure on the one hand, and the stability of the harmonization on the other hand (\ssref{analytical}).
Subsequently, the heuristic control design method is proposed (\ssref{tuning}).
The proposed approach is largely illustrated for the case of force excitation.
The transfer to base excitation is discussed in \ssref{baseExcitation}.

\subsection{Analytical discussion of stability\label{sec:analytical}}
Consider the governing equations of structure and exciter:
\ea{
\ddot \eta_\ell + 2D_\ell\omega_\ell \dot\eta_\ell + \omega_\ell^2\eta_\ell + d_\ell &=& \varphi_{\mrm{ex},\ell}f \quad \ell=1,\ldots,M\fk \label{eq:structure}\\
f &=& \frac{G}{R}u - m_{\mrm{ex}}\left(\ddot q_{\mrm{ex}} + 2D_{\mrm{ex}}\omega_{\mrm{ex}}\dot q_{\mrm{ex}} + \omega_{\mrm{ex}}^2 q_{\mrm{ex}}\right) + d_{\mrm{ex}}\fk \label{eq:exciter}\\
q_{\mrm{ex}} &=& \sum_{\ell=1}^M \varphi_{\mrm{ex},\ell}\eta_\ell\fp \label{eq:qex}
}
Herein, the differential equation of motion of the structure is written in terms of the modal coordinates, $\eta_\ell$, of the underlying linear, conservative, time-invariant system (\eref{structure}).
$D_\ell$, $\omega_\ell$ and $\varphi_{\mrm{ex},\ell}$ are the modal damping ratio, angular frequency and mass-normalized deflection shape at the drive point.
The common model of an electro-dynamic exciter, see \eg \cite{Tomlinson.1979,McConnell.2008}, is used (\eref{exciter}) with the force generating constant $G>0$ and coil resistance $R>0$.
The moving mass of the exciter is $m_{\mrm{ex}}>0$.
The mechanical stiffness, and the effective damping composed of mechanical dissipation and back-electromotive force are described in terms of exciter modal frequency $\omega_{\mrm{ex}}$ and damping ratio $D_{\mrm{ex}}$.
It is assumed that the stinger is rigid, so that the shaker armature is regarded as directly attached to the structure at the drive point, which is expressed in \eref{qex}.
$d_\ell$ and $d_{\mrm{ex}}$ represent deviations from the explicitly stated terms, in particular, nonlinear behavior.
\\
To gain deep qualitative understanding, some (further) simplifications are needed.
First, we assume a $T$-periodic response, where $T=2\pi/\Omega$, with slowly varying higher harmonic Fourier coefficients.
Second, we assume that the adaptive filter acts on a much faster time scale, so that we can idealize the estimate $\tilde{F}_h=F_h$ in \eref{errh}.
Third, we assume that the fundamental harmonic control acts on a slower time scale, so that we can assume $\Omega$ (as well as the fundamental harmonics of $u$, $\eta_\ell$ and $f$) as constant.
Each harmonic $h$ can either be in resonance with a particular mode, or not.
We focus on the resonant case, which is more critical from a control perspective, and briefly discuss the non-resonant case at the end.
We assume that the $h$-th harmonic, with $h\geq 2$, is near resonance with mode $\ell$,
\ea{h\Omega\approx \omega_\ell\fk\label{eq:resonance}
}
and that this is a well-separated modal frequency, so that $h\Omega\approx \omega_n$ does not hold for all $n\in\lbrace 1,\ldots,M\rbrace\backslash\lbrace \ell\rbrace$.
It is further assumed that the resonant mode dominates the respective harmonic contribution, $Q_{\mrm{ex},h}\approx \varphi_{\mrm{ex},\ell}\hat{\eta}_{\ell,h}$, where $Q_{\mrm{ex},h}$ and $\hat{\eta}_{\ell,h}$ are the $h$-th complex Fourier coefficients of $q_{\mrm{ex}}$ and $\eta_\ell$, respectively.
\erefs{structure}-\erefo{qex} can then be cast into the frequency domain,
\ea{
S_{\ell}\left(h\Omega\right)\,\hat{\eta}_{\ell,h} + D_{\ell,h} &=& \varphi_{\mrm{ex},\ell}F_h\fk \label{eq:structureFD}\\
F_h &=& \frac{G}{R}U_h - m_{\mrm{ex}}\varphi_{\mrm{ex},\ell}\,S_{\mrm{e}}\left(h\Omega\right)\,\hat{\eta}_{\ell,h} + D_{\mrm{ex},h}\fk \label{eq:exciterFD}
}
where $S_\ell(\Omega) = -\Omega^2+2D_\ell\omega_\ell\ii\Omega + \omega_\ell^2$ and $S_{\mrm e}(\Omega)=-\Omega^2+2D_{\mrm{ex}}\omega_{\mrm{ex}}\ii\Omega+\omega_{\mrm{ex}}^2$.
After some algebraic manipulations of \erefs{structureFD}-\erefo{exciterFD} and \erefs{errh}-\erefo{Ihdot}, one obtains
\ea{
\dot I_h = - \frac{k_{\mrm{i}}\frac GR I_h + D_h}{1 + k_{\mrm p}\frac GR + Z_{\mrm{e},\ell}\left(h\Omega\right)} \fk \label{eq:IhdotSimplified}
}
where $D_h=D_{\mrm{ex},h} + Z_{\mrm{e},\ell} D_{\ell,h}$, $Z_{\mrm{e},\ell}\left(h\Omega\right)=\mu_{\mrm{ex},\ell}{S_{\mrm{e}}\left(h\Omega\right)}/{S_\ell\left(h\Omega\right)}$, and $\mu_{\mrm{ex},\ell} = m_{\mrm{ex}}\varphi_{\mrm{ex},\ell}^2$.
\\
For a successful suppression, we want the steady-state control error vanish, \ie, $F_h=0$.
Since $\dot I_h=-\tilde{F_h}$ (\eref{Ihdot}), and we assumed $\tilde{F}_h=F_h$, this means that we want to reach the fixed point $\dot I_h=0$ of \eref{IhdotSimplified}.
At this fixed point, we have $U_h=k_{\mrm i} I_h = -D_h R/G$ (\cf \eref{IhdotSimplified}, \eref{Uh}).
\\
So far, we did not make any assumption on $d_{\mrm{ex}}$ and $d_{\ell}$, or the resulting Fourier coefficient $D_h$.
In general, the Fourier coefficients $U_h$ determine $u$, which affects $f$ and $\eta_\ell$.
Therefore, $d_{\mrm{ex}}$, $d_{\ell}$ and, thus, $D_h$, may depend on all $I_n=U_n/k_{\mrm i}$ with $n\neq h$ in a complicated and nonlinear way.
If we assume, for simplicity, that the system is linear and $D_h$ is imposed, the fixed point of \eref{IhdotSimplified} is asymptotically stable if and only if
\ea{
\real{ \frac{k_{\mrm{i}}}{1 + k_{\mrm p}\frac GR + Z_{\mrm{e},\ell}\left(h\Omega\right)} } > 0\fp \label{eq:stability}
}
In the stable case, the magnitude of the real part also determines how quickly we approach the fixed point.
Let's discuss the technical implications of this mathematical result in the following.
\\
It is important to note that the magnitude and the sign of the real part of $Z_{\mrm{e},\ell}\left(h\Omega\right)$ varies rapidly with the fundamental frequency $\Omega$ in the considered resonant case (\eref{resonance}).
Suppose we drive the exciter far above its resonance so that we have $S_{\mrm e}(h\Omega)\approx -(h\Omega)^2$.
Typically, we are interested in testing the nonlinear structure in a sufficiently wide range around resonance.
Before resonance, $h\Omega<\omega_\ell$, $S_\ell(h\Omega)$ assumes a positive real part, at resonance, $h\Omega=\omega_\ell$, a zero real part, and above resonance, $h\Omega>\omega_\ell$, a negative real part.
As a consequence, the sign of the real part in \eref{stability}, and hence the stability, may change when $\Omega$ is varied around resonance, for otherwise constant parameters.
To ensure stability throughout the frequency range of interest in such a case, in principle, one would have to switch the sign of $k_{\mrm i}$ at precisely the right moment.
This seems rather impracticable.
Therefore, such a case should be avoided.
To achieve this, one should make $k_{\mrm p}$ sufficiently large, and make sure that the magnitude of $Z_{\mrm{e},\ell}\left(h\Omega\right)$ remains sufficiently small.
For the practically relevant case of light damping, $0<D_\ell\ll 1$, the largest magnitude is reached at resonance, $h\Omega=\omega_\ell$.
There, we have $S_\ell=2D_\ell\omega_\ell^2\ii$, and thus $\|Z_{\mrm{e},\ell}\|=\|\mu_{\mrm{ex},\ell}/(2D_\ell)\|$.
To reach a sufficiently small magnitude $\|Z_{\mrm{e},\ell}\|$, we thus have to select exciter and drive point in such a way that $\mu_{\mrm{ex},\ell}$ is sufficiently small compared to the damping ratio $D_\ell$.
In other words, stable control is more difficult to achieve in the very lightly damped case, where also high $\|\hat\eta_{\ell, h}\|$ is expected.
It is useful to note that the fraction $\mu_{\mrm{ex},\ell}/(2D_\ell)$ is a key parameter also for other forms of exciter-structure interaction, including the well-known force dropout near resonance \cite{Krack.2021}.
An important finding is also that a proportional gain increases robustness of the controller, but also slows it down.
Further, the higher the integral gain, $k_{\mrm i}$, the faster the harmonization.
In practice, the gains $k_{\mrm p}$ and $k_{\mrm i}$ cannot be set arbitrarily large.
Important limitations are the stability loss due to magnification of noise, which was not included in the simple analysis above, possible unstable interaction with the fundamental harmonic controller or the adaptive filter, and the input voltage limit of the (power amplifier of the) exciter.

\subsection{Heuristic control design\label{sec:tuning}} 
The key parameters of the harmonization module are the proportional and integral gain, $k_{\mrm p}$ and $k_{\mrm i}$.
Recall that it is proposed to set the truncation order $H$ equal to that of the adaptive filter.
Also, the cutoff frequency $\omLP$ of the adaptive filter must be set.
The purpose of this section is the design of those control parameters.
\\
As explained in the beginning of this section, a nonlinear-model-based design seems useless (since a valid nonlinear model is not available prior to the test), so that a heuristic control design is proposed instead.
What can be assumed to be available is the target resonance condition, and a good estimate of the modal frequencies of the associated linear modes.
By \emph{target resonance condition}, the relation between the frequency of the sinusoidal excitation and the modal frequencies is meant.
For instance, the fundamental excitation frequency can be near a primary or a secondary resonance with a modal frequency, and there can be internal resonances, \eg, another modal frequency can be at an integer multiple of the externally resonant one.
\\
In general, it seems impractical to adjust the control parameters during the test; the aim is to obtain a set of fixed parameters which ensures robust control performance in the relevant range of excitation frequency and level.
It is proposed \emph{to run a test without the harmonization module first}.
This first test can be a relatively quick one, for instance a coarse stepped or moderately slow swept sine test, without or with control of the excitation level.
The goal is to \emph{determine a representative point}, in terms of excitation frequency and level, which can be used for the subsequently proposed tuning scheme.
The tuning is typically more difficult at lower signal levels, due to the more pronounced effect of noise.
On the other hand, the higher harmonics must not be negligible, which means that the vibration level must be high enough to activate sufficiently nonlinear behavior.
A practical choice is the upper or lower bound of the considered frequency window, in conjunction with the lowest tested excitation level showing non-negligible nonlinear behavior.
\\
An overview of the proposed heuristic scheme is given here:
\begin{algorithmic}[1]
\LComment{TEST WITHOUT HARMONIZATION; TUNING OF ADAPTIVE FILTER; SETTING OF TRUNCATION ORDER}
\State \msp Run a test without the harmonization module.
\State \msp Determine a representative excitation frequency and level.
\State \msp Select $H$ under consideration of target resonance condition and sampling rate.
\State \msp Select maximum $\omLP$ so that fluctuations in $\tilde F_h$ are acceptable at the representative point.
\LComment{TUNING OF GAINS}
\State \msp Set $k_{\mrm i}=0$ and successively increase $k_{\mrm p}$.
\State \msp Select $k_{\mrm p}$ sufficiently distant from stability boundary $k_{\mrm p}^{\mrm{crit}}$.
\State \msp Maintain selected $k_{\mrm p}$ and successively increase $k_{\mrm i}$.
\State \msp Select $k_{\mrm i}$ sufficiently distant from stability boundary $k_{\mrm i}^{\mrm{crit}}$.
\end{algorithmic}
The selection of the individual parameters is described in the following.

\subsubsection*{Adaptive filter cutoff frequency $\omLP$}
Concerning the cutoff frequency $\omLP$ of the adaptive filter, the tuning proposed in \cite{Hippold.2024} is adopted.
It is important to understand that frequency components different from those retained in the adaptive filter ($0$, $\Omega$, $2\Omega$, \ldots, $H\Omega$), including broadband noise, lead to fluctuations in the estimated Fourier coefficients (\cf \eref{AF}).
Those fluctuations limit the achievable control quality.
The higher $\omLP$, the higher the fluctuations.
However, $\omLP$ also defines the time scale of the control; \ie, faster control can be achieved with higher $\omLP$.
Thus, a good compromise between quality under noise and speed must be found.
\\
To find a suitable value for $\omLP$, it is proposed to do an \emph{open-loop-test} at the representative frequency and level, identified as described above.
A range of adaptive filters is applied and that with the highest $\omLP$ is selected, which leads to fluctuations of the Fourier coefficients smaller than a given tolerance.
Note that this step can be run online in parallel, or even applied offline to the acquired excitation and response signals.
The tolerance should be significantly smaller than the target tolerance of the harmonization.
We recommend that if the control target is $\|\tilde F_h\|/\|\tilde F_1\| < \errtol$, then the fluctuations should be smaller than $\errtol/2$.
For wider ranges of excitation frequency and level, requiring $\errtol/10$ or less fluctuations might be useful.
It is proposed to consider values of $\omLP$ in the range $1/100<\omLP/\omega_{1}<1$, where $\omega_{1}$ is the linear modal frequency of the resonant mode.
Higher values are likely to lead to a non-robust controller, while lower values are likely to yield an impracticably long test duration.
Thus, if the phase fluctuations are still unacceptable at $\omLP/\omega_{\mrm{lin}}=1/100$, one should consider improving the signal-to-noise ratio.
This can be achieved by modifying the instrumentation (to reduce noise), and/or to start the vibration test at a higher initial level (to increase the signal strength).

\subsubsection*{Truncation order $H$}
The suppressed higher harmonics should at least span the resonant ones.
For instance, if a $1:3$ internal resonance is present, the second and third harmonic should at least be suppressed; \ie, $H\geq3$.
If the relevant sources of nonlinear behavior are known to be smooth, $H$ probably does not have to be much larger.
Running vibration tests with successively increased $H$ can be useful to gain confidence in the robustness of the results to the selected truncation order.
In any case, the truncation order must be sufficiently small to avoid aliasing for the given sampling rate; \ie, $H<2\pi \nu_{\mrm s}/\Omega_{\max}$, where $\nu_s$ is the sampling frequency and $\Omega_{\max}$ is the highest fundamental (angular) excitation frequency.

\subsubsection*{Gains $k_{\mrm p}$ and $k_{\mrm i}$\label{sec:gainTuning}}
Once the adaptive filter and truncation order have been set, the gains are to be selected.
Again, these parameters are tuned for the representative point, identified as described above.
Recall that it is proposed to use the same gains for all harmonics.
It is proposed to keep the fundamental harmonic controller active during the entire tuning procedure.
\\
The proposed heuristic scheme starts with $k_{\mrm i}=0$, and successively increasing $k_{\mrm p}$.
It is expected that the control error $\|\tilde F_h\|$ decreases at first, until pronounced oscillations of the Fourier coefficients start to occur at some $k_{\mrm p}^{\mrm{crit}}$, indicating the stability boundary.
It is proposed to select a $k_{\mrm p}$ with some margin from that critical value, say $k_{\mrm p} = k_{\mrm p}^{\mrm{crit}}/2$.
Next, while keeping this $k_{\mrm p}$, the analog scheme is applied to find a suitable $k_{\mrm i}$.
\\
It is important to monitor the evolution of the individual Fourier coefficients $\tilde F_h$ over time.
This applies, in particular, when moving away from the representative point for which the harmonization has been designed.
As stated before, harmonic-specific gains can easily be implemented, but this makes the gain tuning much more complicated, and did not increase significantly the (already high) control performance, throughout the virtual and real experiments investigated so far.

\subsection{From force to base excitation\label{sec:baseExcitation}}
Base excitation is actually used in the real experiment (\sref{exp_vali}).
In this case, the applied excitation is not the force $f$ but the base motion.
It is then useful to split the absolute motion, $\mm b q_{\mrm b}+\mm q$, into base displacement $q_{\mrm b}$ and elastic deformation $\mm q$ relative to the base.
The vector $\mm b$ is Boolean in suitable coordinates, with entry 1 if the respective coordinate is aligned with the base motion, and entry 0 if it is orthogonal.
The modal forcing on the right-hand side of \eref{structure} becomes $-\mm\phi^{\mrm T}_\ell \mm M\mm b \ddot{q}_{\mrm b}$, $q_{\mrm{ex}}$ must be replaced by $q_{\mrm b}$ in \eref{exciter}, and inertia feedback forces related to the structural vibrations have to be added, $\mm b^{\mrm{T}}\mm M \mm\phi_n \ddot{\eta}_n$.
For details on the modeling of base excitation in this context, we refer to \cite{Muller.2022}.
Most of the above discussion applies also to base excitation, including the finding that a low fraction $\mu_{\mrm{ex},\ell}/(2D_\ell)$ facilitates the design of a good controller.
One exception is the relevant mass ratio, which becomes $\mu_{\mrm{ex},\ell} = (\mm b^{\mrm{T}}\mm M \mm\phi_\ell)^2/m_{\mrm{ex}}$ in the case of base excitation.
This implies that a large moving mass $m_{\mrm{ex}}$ is desired to achieve a low $\mu_{\mrm{ex},\ell}$, in complete contrast to case of force excitation.

\section{Virtual experiment: validation\label{sec:num_vali}}
The benefit of a virtual experiment is that a perfectly mono-harmonic forcing can be simply imposed in the simulation, to obtain an ideal reference.
Thus, the effect of the residual higher harmonic content on the nonlinear response of the structure can be precisely quantified.
The example is illustrated in \fref{virt_setup} and was adopted from Shaw \etal \cite{Shaw.2016}.
It consists of a beam with a cubic spring, harmonically driven near the primary resonance of the lowest-frequency mode, which is near a 1:3 internal resonance with the second mode.
This example was selected because, first, identified models of structure and matching exciter are available from real experiments, and, second, the structure exhibits strongly nonlinear behavior.
In particular, the frequency response features turning points of the main branch and an isolated branch, for appropriate forcing level.
\begin{figure}[h!]
    \centering
    \includegraphics{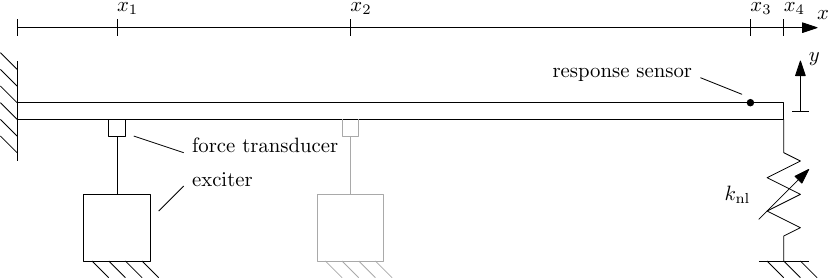}
    \caption{Virtual experiment: Schematic illustration \cite{Shaw.2016}.}
    \label{fig:virt_setup}
\end{figure}
\begin{table}[h!]
    \centering
    \caption{Virtual experiment: Modal parameters of the structure \cite{Shaw.2016}.}
    \label{tab:modal_para}
    \begin{tabular}{ccccccc}
    \hline
       mode $n$ & $\omega_{n,\mathrm{lin}}$ in rad/s & $D_{n,\mathrm{lin}}$ in \% & $\phi_{1,n}$ in $1/\sqrt{\mrm{kg}}$ & $\phi_{2,n}$ in $1/\sqrt{\mrm{kg}}$ & $\phi_{3,n}$ in $1/\sqrt{\mrm{kg}}$ & $\phi_{4,n}$ in $1/\sqrt{\mrm{kg}}$ \\
       \hline\hline
       1  & 55.92 & 1.00 & 0.125 & 1.35 & 5.13 & 5.34 \\
       2  & 199.18 & 1.00 & -0.575 & -3.86 & 3.8 & 4.67 \\
       \hline
    \end{tabular}
\end{table}
\\
The dynamics of the structure and the exciter are governed by the equations:
\ea{
\ddot{\eta}_\ell + 2D_\ell\omega_\ell\dot{\eta}_\ell +\omega_\ell^2\eta_\ell + d_{\ell}(\eta_1,\eta_2) &=& \phi_{\mrm{ex},\ell} f \qquad \ell=1,2\fk \label{eq:structureShaw}\\
f &=& \frac{G}{R} u - \mex \left( \ddot{q}_{\mrm{ex}} + 2\Dex\omex \dot{q}_{\mrm{ex}} + \omex^2 q_{\mrm{ex}} \right) \fk \label{eq:exciterShaw} \\
q_{\mrm{ex}} &=& \sum_{n=1}^2 \phi_{\mrm{ex},n}\eta_n\fk \label{eq:qexShaw} \\
d_\ell &=& \phi_{\mrm{nl},\ell} k_\mrm{nl} \left(\sum_{n=1}^2 \phi_{\mrm{nl},n} \eta_n \right)^3\fp \label{eq:dellShaw}
}
This is a special form of the more general equations \erefo{structure}-\erefo{qex} considered in the theoretical part.
More specifically, the structure has been truncated to $M=2$ modes, and the distortion terms are $d_\ell$ as defined in \eref{dellShaw}, and $d_{\mrm{ex}}=0$.
Two modes were found sufficient to reproduce the strongly nonlinear behavior, including the modal interactions and the formed isolated branch observed in the real experiment of Shaw \etal \cite{Shaw.2016}.
The modal parameters are listed in \tref{modal_para}.
The parameter $k_\mrm{nl}$ of the cubic spring was set to \SI{2.517e6}{\newton\per\cubic\meter} \cite{Shaw.2016}.
The cubic spring was attached at the free end ($x_4$ in \fref{virt_setup}), so that $\phi_{\mrm{nl},\ell}=\phi_{4,\ell}$.
\begin{table}[h!]
    \centering
    \caption{Virtual experiment: Exciter parameters.}
    \label{tab:exc_para}
    \begin{tabular}{ccc}
    \hline
        parameter & symbol &  value\\
        \hline\hline
         armature mass & $m_\mathrm{ex}$ & \SI{0.057}{\kilo\gram} \\
        coil resistance & $R$ & \SI{2}{\ohm} \\
        force constant & $G$ & \SI{6.78}{\newton\per\ampere} \\
        exciter natural frequency & $\omex$ & \SI{417.4}{\per\second} \\
        exciter damping ratio & $\Dex$ & 0.935 \\
        \hline
    \end{tabular}
\end{table}
\\
The exciter can be attached either at the location $x_1$ or $x_2$ (\fref{virt_setup}), so that $\phi_{\mrm{ex},\ell}$ has to be set to $\phi_{1,\ell}$ or $\phi_{2,\ell}$, respectively.
For the exciter used in the real experiments of \cite{Shaw.2016}, no identified model was available.
Instead, we used the parameters of a \textsc{Brüel \& Kj\ae r} Type 4809 vibration exciter, identified in our laboratory, and listed in \tref{exc_para}.
It has been verified that frequency and force range of this exciter are suitable for the considered vibration test.

\subsection{Selection of drive point}
In this subsection, we illustrate the importance of an appropriate selection of the drive point.
To this end, we revisit the analytical stability criterion derived in \eref{stability}: 
For positive integral gain, $k_{\mrm i}>0$, asymptotic stability requires a positive real part of the complex transfer function $1/(1+\bar{k}_{\mrm p} + Z_{\mrm e})$.
Herein, $\bar{k}_{\mrm p}=k_{\mrm p}G/R$ is the dimensionless proportional gain, and we have generalized $Z_{\mrm e}$ from the single-resonant-mode approximation $Z_{\mrm{e},\ell}$ in \eref{stability} to $Z_{\mrm{e}}=\sum_{\ell=1}^{M} Z_{\mrm{e},\ell}$, in order to account for contributions of off-resonant modes.
The real part of the above stated transfer function is shown in \fref{TF}(a) and (b) for drive point $x_1$ and $x_2$, respectively.
The parameters of the underlying linear model of the structure and the exciter, as listed in \trefs{modal_para}-\trefo{exc_para}, have been used.
\begin{figure}[h!]
    \centering
            \begin{subfigure}{0.49\textwidth}
        \centering
        \includegraphics{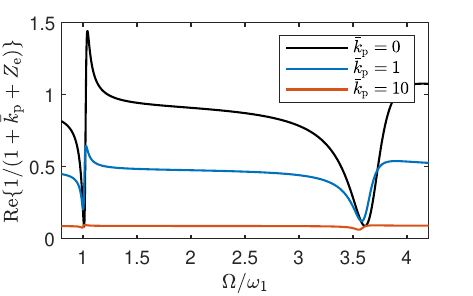}
        \caption{}
    \end{subfigure}
    \hfill
    \begin{subfigure}{0.49\textwidth}
        \centering
        \includegraphics{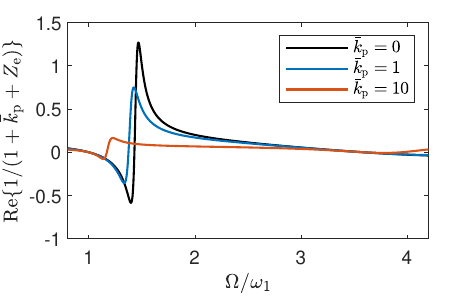}
        \caption{}
    \end{subfigure}
    \caption{Real part of complex transfer function in \eref{stability} for (a) drive point $x_1$, and (b) drive point $x_2$.}
    \label{fig:TF}
\end{figure}
\\
Recall that the transfer function is to be evaluated at $h\Omega$, where $\Omega\approx\omega_1$ in the present case of a primary resonance with the fundamental mode, and $h=2,3,\ldots,H$.
For drive point $x_2$, the real part has a negative zero crossing near $\Omega/\omega_1=3$, so that instability of the harmonization loop is expected.
The situation is not much improved with the proportional gain.
For drive point $x_1$, in contrast, the real part remains positive, so that stable control is expected.
The proportional gain reduces the sensitivity with respect to the excitation frequency.
The expected better control performance for this drive point is attributed to the lower mass ratio.
In this example, the mass ratio $\mu_{\mrm{ex},\ell}=m_{\mrm{ex}}\phi_{\mrm{ex},\ell}^2$ reduces from $10^{-1}$ to $<10^{-3}$ for mode 1, and from $0.85$ to $<0.02$ for mode 2, when moving the drive point from $x_2$ to $x_1$.
This analysis shows how important the selection of the drive point is for vibration control tasks.
A further reduction of $\mu_{\mrm{ex},\ell}$ could be achieved by moving the drive point even closer to the clamping.
However, at some point, the selected exciter will not be able to provide sufficient forcing to drive the system into the desired vibration levels.
In the subsequent study, we use only drive point $x_1$.

\subsection{Control design: gain selection}
A conventional fundamental harmonic control was used to maintain the applied force level constant.
To this end, fundamental Fourier coefficient $\tilde F_1$ was estimated with the adaptive filter used also for the harmonization module.
The control error was specified as $\|\tilde F_1\|-\hat F_1$ to achieve a target value of $\hat F_1\in\mathbb R_{>0}$.
A simple integral controller with gain $\SI{0.1}{\volt\per\newton\per\second}$ was used.
With this, the fundamental harmonic controller is slow compared to the harmonization.
No systematic tuning was carried out for the fundamental harmonic control.
\begin{table}[h!]
    \centering
    \caption{Virtual experiment: Selected controller parameters.}
    \label{tab:control_virt}
    \begin{tabular}{ll}
    \hline
        parameter & value \\
        \hline\hline
        truncation order & $H=7$ \\
        adaptive filter cutoff frequency & $\omLP = \omega_{1} / 10$ \\
        proportional gain & $k_{\mrm{p}}G/R = 3$\\
        integral gain & $k_{\mrm{i}}G/R/\omLP = 2$\\
        \hline
    \end{tabular}
\end{table}
\\
The real experiment in \cite{Shaw.2016} was limited to the suppression of the third harmonic.
Due to the symmetry of the nonlinear term, no even harmonics are generated for odd harmonic input.
To test the robustness of the proposed harmonization, we set $H=7$ and apply the feedback control also to the even harmonics.
The magnitude of force harmonics of order 8, 9 and higher was deemed negligible.
No noise source was considered, so that the proposed selection of $\omLP$ could not be applied.
Instead, we use $\omLP = \omega_{1} / 10$, which is a very typical value in our experience gained so far.
\\
The gains were tuned as proposed in \sref{gainTuning}.
As representative excitation frequency and level, $\Omega=\omega_1$ and $\hat{F}_1 = \SI{2}{\newton}$ were used, respectively.
For each set of tested gains $k_{\mrm p}$, $k_{\mrm i}$, the simulation was started with only the fundamental harmonic control active, and it was waited until this had settled, before activating the harmonization at $t=0$.
The left column of \fref{virtualTuning} illustrates the tuning of the proportional gain $k_{\mrm p}$.
\fref{virtualTuning}e shows the onset of oscillations of the Fourier coefficients $\tilde F_h$.
As proposed, $k_{\mrm p}$ was set to $50\%$ of this value.
The simulation with the resulting $k_{\mrm p}$ is shown in \fref{virtualTuning}g.
The subsequent tuning of the integral gain $k_{\mrm i}$ is shown in the right column of \fref{virtualTuning}.
Oscillations of the Fourier coefficients $\tilde F_h$ now arise in \fref{virtualTuning}f.
As proposed, $k_{\mrm i}$ was set to $50\%$ of this value.
The simulation with the finally selected set of gains is shown in \fref{virtualTuning}h.
\begin{figure}[!ht]
    \centering
    \includegraphics[]{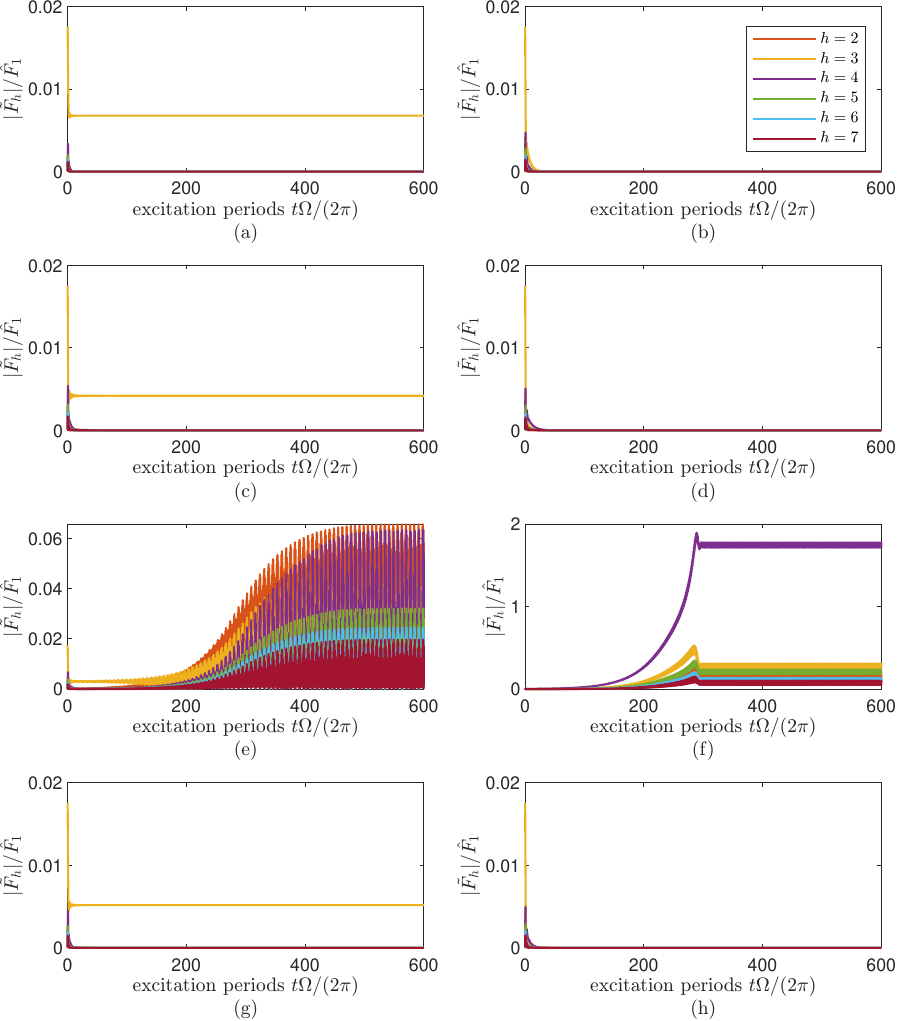}
    \caption{Tuning of the controller gains in the virtual experiment. Left column: tuning of the proportional gain ($\ki = 0$), (a) $\kpnd = 2$, (c) $\kpnd = 4$, (e) $\kpnd = 6$, and (g) selected value $\kpnd = 3$. Right column: tuning of the integral gain ($\kpnd = 3$), (b) $\kind = 1$, (d) $\kind = 2.5$, (f) $\kind = 4$, and (h) selected value $\kind = 2$. $\kpnd = \kp G/R$, $\kind = \ki G/R/\omLP$}
    \label{fig:virtualTuning}
\end{figure}
\\

\subsection{Harmonized virtual tests vs. numerical reference}
Throughout this section, the focus is placed on the frequency range around the primary resonance with the fundamental mode, and the target force level was set to $\hat F_1 = 2~\mrm N$.
For this force level, the structure shows strongly nonlinear behavior, including an isolated frequency response branch.
The results obtained for forward frequency stepping along the main branch are shown in \frefs{virt_exp_main}, and those for the isolated branch in \frefo{virt_exp_isola}.
As response quantity, the displacement at the location $x_3$ (\fref{virt_setup}), $q=\sum_{\ell=1}^2\phi_{3,\ell}\eta_\ell$, is used.
\begin{figure}[p!]
    \centering
    \includegraphics[]{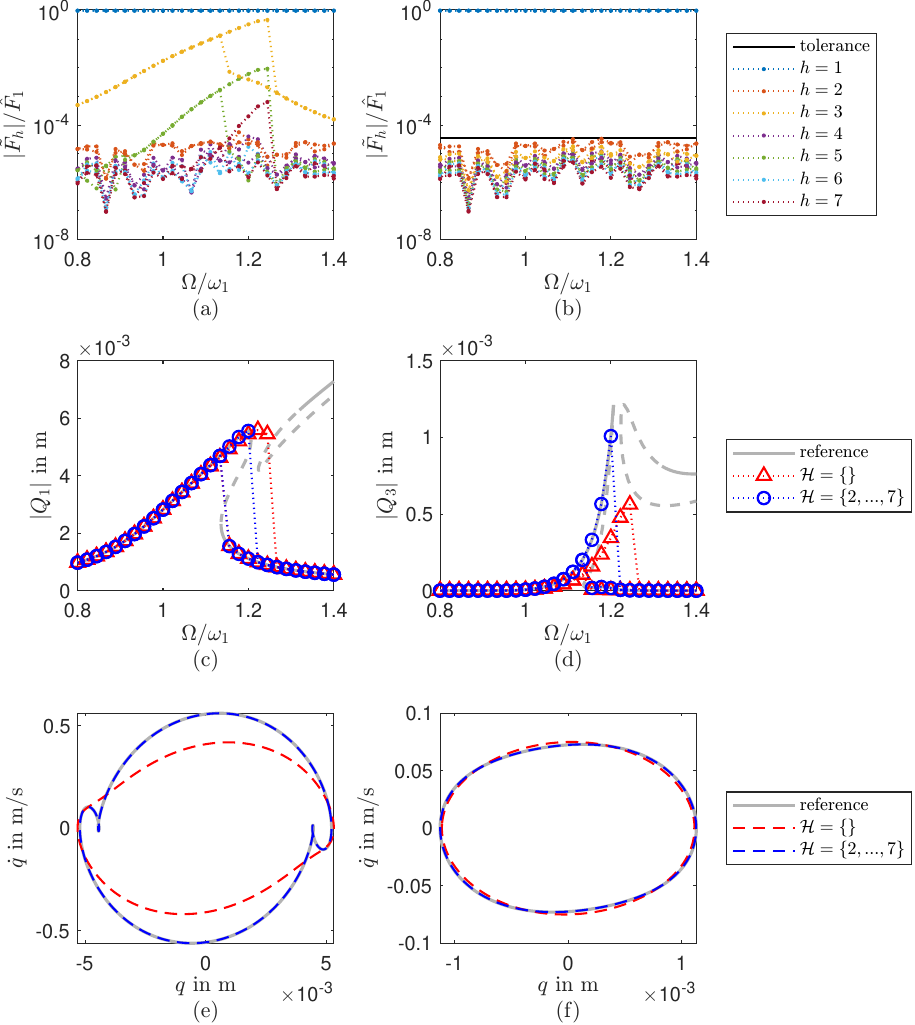}
    \caption{Virtual experiment: Stepped sine results of \emph{main branch}:
    Fourier coefficients of applied force (a) without ($\mathcal{H}=\lbrace\rbrace$) and (b) with active harmonization ($\mathcal H=\lbrace 2,\ldots,7\rbrace$);
    magnitude of (a) fundamental and (b) third Fourier coefficient of response displacement;
    phase projection at $\Omega/\omega_1 = 1.2$ on (e) lower and (f) upper part of main branch.
    Asymptotically stable response regimes are indicated by solid, and unstable by dashed parts of the reference curve.
    }
    \label{fig:virt_exp_main}
\end{figure}
\begin{figure}[p!]
    \centering
    \includegraphics{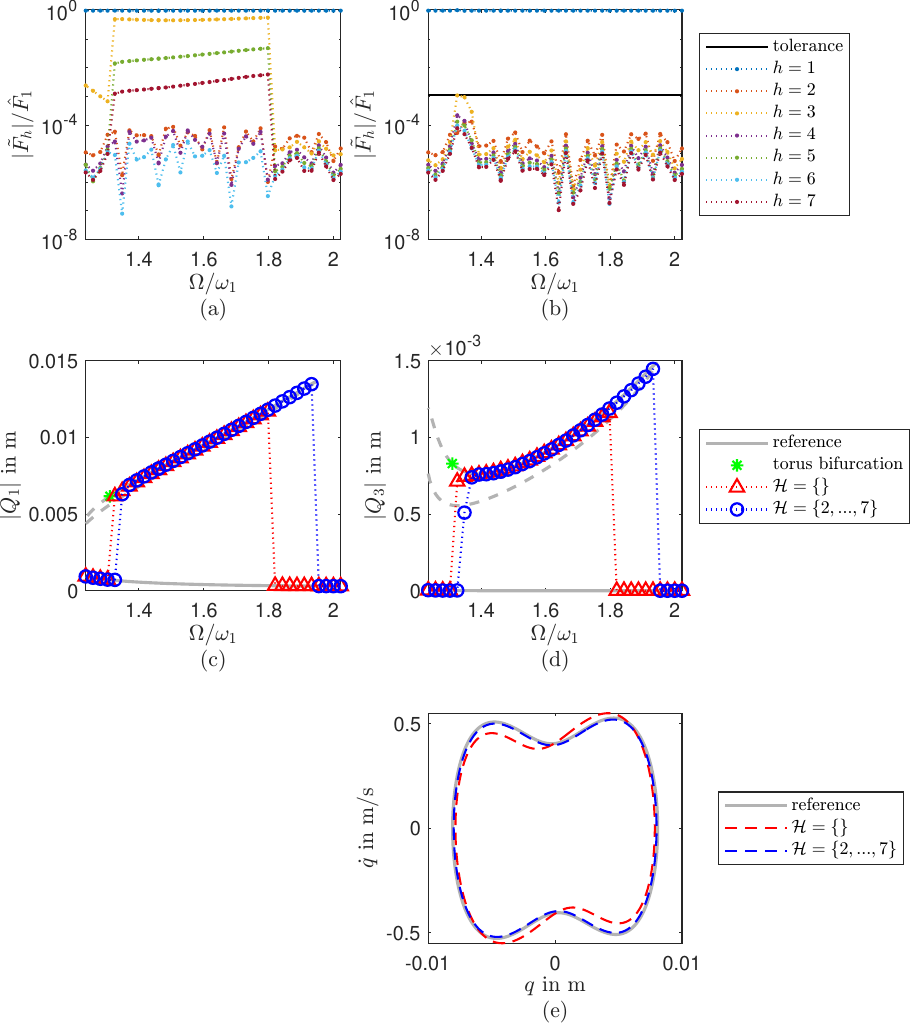}
    \caption{Virtual experiment: Stepped sine results of \emph{isolated branch}:
    Fourier coefficients of applied force (a) without ($\mathcal{H}=\lbrace\rbrace$) and (b) with active harmonization ($\mathcal H=\lbrace 2,\ldots,7\rbrace$);
    magnitude of (c) fundamental and (d) third Fourier coefficient of response displacement;
    (e) phase projection at $\Omega/\omega_1 = 1.4$ on stable part of isolated branch.
    Asymptotically stable response regimes are indicated by solid, and unstable by dashed parts of the reference curve.
    }
    \label{fig:virt_exp_isola}
\end{figure}
\\
For the virtual stepped sine tests, numerical forward time step integration was employed to solve the system of ordinary differential equations governing the behavior of structure, exciter, fundamental harmonic controller and harmonization module.
The explicit fifth-order Runge-Kutta Dormand-Prince scheme was used with automatic step size adjustment and a maximum time step of $10^{-3}~\mrm{s}$.
This ensures a minimum of about 30 time steps per natural period of the second mode, and it was ensured that a further reduction of the maximum time step has no significant effect on the depicted results.
At the first frequency point, homogeneous initial conditions were specified.
To smoothly transition from one frequency to the next, half-cosine ramps with a duration of 10 periods, $2\pi/\omega_1$, of the lowest-frequency mode were used.
After each ramp, the frequency was maintained for 600 periods.
The last 300 excitation periods of those hold phases were used to estimate Fourier coefficients of the response and the applied forcing via fast Fourier transform (FFT).
It was verified that the controllers had settled sufficiently in the post-processed time frame, and that this time frame was long enough to obtain meaningful results.
As alternative to this conventional procedure involving hold phases of prescribed duration, and using an FFT-based post-processing of the steady state, it seems feasible and useful to implement an automatic settling detection in future work, and directly use the Fourier coefficients estimated by the adaptive filter, as proposed in \cite{Hippold.2024}.
\\
The \emph{reference} was obtained using the shooting method implemented in \textsc{NLvib} \cite{Krack.2019}.
This relies on the constant-average-acceleration Newmark scheme for time step integration.
A number of one thousand time levels per excitation period was found sufficient according to a convergence study.
For the numerical reference, mono-harmonic forcing, $\hat F_1\cos\left(\Omega t\right)$, was directly imposed at the drive point.
The asymptotic stability of the periodic orbits was inferred from the Floquet multipliers derived from the monodromy matrix obtained as by-product of shooting.
\\
To reach the isolated branch using the reference method, the phase-resonant backbone curve was numerically continued, where the force level is treated as unknown.
This way, three points are reached for which the force level equals the target value.
The upper two points intersect with the isolated branch, so that conventional path continuation can be initialized from these points to obtain the isolated branch.
To reach the isolated branch in the virtual experiment, the procedure used in \cite{Shaw.2016} was adopted.
Here, the stepped sine test is run forward along the main branch until shortly before the peak.
Then, the excitation frequency and the fundamental input voltage $U_1$ are suddenly increased by a suitable value found by trial and error.
Due to the active fundamental controller, the applied force level settles back to its target value.
The harmonization module was permanently active to verify its robustness to such sudden events.
\\
As can be seen in \fref{virt_exp_main}a-b, \fref{virt_exp_isola}a-b, the fundamental harmonic controller performed well, \ie, $\|\tilde F_1\|/\hat F_1\approx 1$, throughout the stepped sine tests.
Without harmonization, pronounced higher harmonics occur, where the third harmonic reaches about $50\%$ (\fref{virt_exp_main}a, \fref{virt_exp_isola}a,).
With harmonization, the higher harmonics are a few orders of magnitude lower, and can be regarded to be in the noise floor.
More specifically, the magnitude of the higher force harmonics remains largely $<0.01\%~\hat F_1$, with the exception of the two lowest frequency levels on the isolated branch (\fref{virt_exp_isola}b).
This is the regime where the periodic response has lost stability via a Torus bifurcation, giving rise to non-periodic behavior.
Evidence of such a bifurcation and non-periodic behavior in this regime was also found in \cite{Shaw.2016}.
Thus, frequency components appear in the response which are not contained in the truncated Fourier series assumed by the adaptive filter (\eref{AF}).
Consequently, the Fourier coefficients estimated by the adaptive filter are distorted by pronounced fluctuations, which limits the quality of the higher harmonic control.
\\
Without harmonization, the stepped sine test leads to a premature jump from the high-level branch, before reaching the maximum response peak.
This applies to the peaks of both, the main branch (\fref{virt_exp_main}c-d) and the isolated branch (\fref{virt_exp_isola}c-d).
Also, the third response harmonic is severely underestimated on the upper part of the main branch (\fref{virt_exp_main}d), which can also be seen in the phase projection (\fref{virt_exp_main}e).
Finally, a phase shift error appears, which can be inferred from the phase projection in \fref{virt_exp_isola}e.
With harmonization, in contrast, the stepped sine test perfectly follows the (asymptotically stable part of the) reference frequency response branches.

\section{Real experiment: robustness and assessment against state of the art\label{sec:exp_vali}}
The benefit of a real experiment is that the robustness of the proposed method to real-world imperfections and noise can be analyzed.
Also, the proposed method is assessed against an iterative state-of-the-art harmonization in this section.
Another difference to the virtual experiment in \sref{num_vali} is that base instead of force excitation was used.
As structure under test, a doubly-clamped beam is considered (\fref{exp_setup}).
The beam is nominally straight, and has a free length of \SI{140}{\milli\meter}, a width of \SI{10}{\milli\meter} and a thickness of \SI{1}{\milli\meter}.
The beam is bolted to a support frame mounted on the armature of a vibration exciter (\textsc{Brüel \& Kj\ae r} Type 4808).
Variants of this test rig were studied also in \cite{Muller.2022,Abeloos.2022,Muller.2023}.
Due to the clamped ends, the beam's bending induces membrane stretching, which leads to pronounced hardening (nonlinear bending-stretching coupling). 
The base motion in the $y$-direction was acquired via acceleration sensors (\textsc{Dytran} 3035B) placed on the clamping blocks, two on each side.
To obtain a scalar value, the average of the four sensor outputs was taken as base acceleration.
The vibration was acquired with a laser-Doppler vibrometer (\textsc{Polytec} PSV400), pointed at one third of the beam's free length (\fref{exp_setup}c), aligned with the $y$-direction.
As response quantity, the $y$-velocity at this location, relative to the base was used.
The base velocity was obtained from the base acceleration in the frequency domain; \ie, using the estimated Fourier coefficients.
A \textsc{dSpace} MicroLabBox was used for data acquisition and the implementation of the controllers, operating at a sampling frequency of \SI{10}{\kilo\hertz}.
\begin{figure}[h!]
\centering
\begin{subfigure}[]{0.45\textwidth}
\centering
\includegraphics[width=\textwidth]{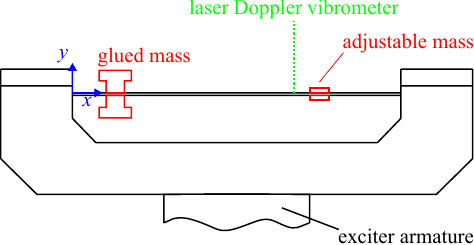}
\caption{}
\end{subfigure}
\hfill
\begin{subfigure}[]{0.45\textwidth}
\centering
\includegraphics[width=\textwidth]{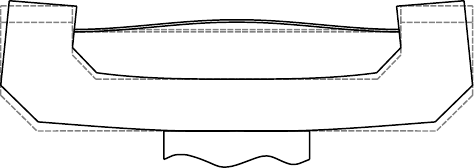}
\caption{}
\end{subfigure}
\begin{subfigure}[]{0.7\textwidth}
\centering
\includegraphics[width=\textwidth]{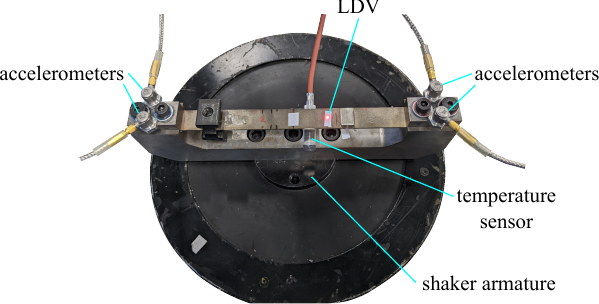}
\caption{}
\end{subfigure}
\caption{Real experiment: (a) arrangement of masses on beam; (b) schematic illustration of frame deformation induced by fundamental bending mode of beam; (d) photo of instrumented test rig.}
\label{fig:exp_setup}
\end{figure}
\\
The setup inherently generates higher excitation harmonics, as explained in the following.
When the beam deforms in a shape similar to the fundamental bending mode, the frame deforms as schematically depicted in \fref{exp_setup}b.
The frame's lowest natural frequency is much higher than that of the beam, so that the frame deforms in good approximation quasi-statically.
The beam's bending induces a clamping deformation in the beam's axial direction.
Due to the frame's geometry, this is associated with a tilting movement of the clamping, and also a displacement in the $y$-direction.
As the membrane stretching occurs at twice the bending frequency, a second harmonic is generated by this mechanism.
\\
The structure under test was designed to respond sensitively to the second harmonic, generated as described above.
To this end, mass elements were attached as shown in \fref{exp_setup}, with the goal to tune the first and second bending mode close to a 1:2 internal resonance.
More specifically, a larger mass (\SI{22}{\gram}) was glued at $x=\SI{18}{\milli\meter}$, and two magnets (total mass \SI{3}{\gram}) were attached at $x=\SI{104}{\milli\meter}$.
Linear modal testing confirmed a natural frequency ratio of $\omega_{2}/\omega_{1} = 2.07$.
Note that sufficiently nonlinear behavior and, hence, sufficiently high vibration levels are needed to enable a pronounced modal interaction.
At higher vibration levels, hardening is expected to cause an appreciable shift of the fundamental modal frequency.
For this reason, a frequency ratio $\omega_2/\omega_1>2$ was intended and the value of $=2.07$ was deemed appropriate.
The arrangement of the attached masses on the beam was designed to break the symmetry of the structure under test.
This was done with the intent to make the second mode excitable under symmetric base motion.

\subsection{Preliminary test without harmonization}
\begin{figure}[h!]
    \centering
    \begin{subfigure}{0.49\textwidth}
        \centering
        \includegraphics{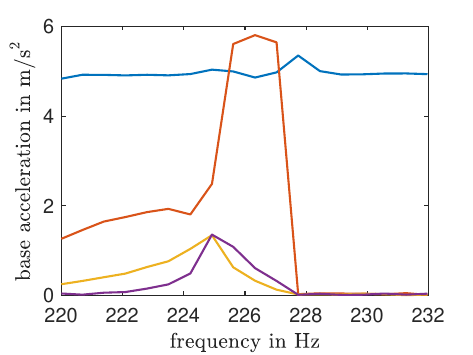}
        \caption{}
        \label{fig:acc_prelim}
    \end{subfigure}
    \hfill
    \begin{subfigure}{0.49\textwidth}
        \centering
        \includegraphics{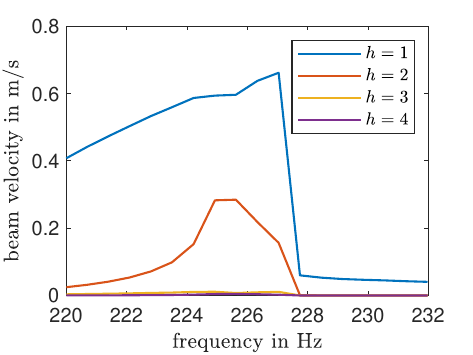}
        \caption{}
        \label{fig:vel_prelim}
    \end{subfigure}
    \caption{Real experiment: Preliminary, coarsely resolved stepped sine test without harmonization; (a) base acceleration (b) response velocity.}
    \label{fig:prelim_test}
\end{figure}
As proposed in \sref{tuning}, a test without harmonization was run first, with relatively coarse frequency spacing and short hold phases.
The target base acceleration was set to $\hat A_{\mrm{b},1}=\SI{5}{\meter\per\square\second}$, where $A_{\mrm{b},h}$ is the $h$-th complex Fourier coefficient of $a_{\mrm b}=\ddot q_{\mrm b}$.
The fundamental harmonic controller was adopted from \cite{Abeloos.2022}.
The results are presented in \fref{prelim_test}.
As expected for hardening, the forward frequency stepping leads to a jump from the high- to the low-level response branch.
Because of the large frequency steps and short hold phases, the fundamental harmonic controller did not settle completely.
This led to a considerable deviation from the set value of $\SI{5}{\meter\per\square\second}$ at some frequency levels.
A pronounced second harmonic appears in the base acceleration, even exceeding the fundamental one in a certain frequency range (\fref{prelim_test}a).
The third and fourth harmonic also contribute to an appreciable extent.
Here and in the following, the \emph{fifth and higher harmonics were deemed negligible}, and are thus not shown.
With the given sampling rate of \SI{10}{\kilo\hertz}, one has about 40 samples per fundamental excitation period; the highest relevant harmonic is thus regarded as well-represented.
A peak of the second response harmonic appears (\fref{prelim_test}b), which indicates a resonant modal interaction.
In contrast, third and fourth response harmonic are negligible, since there is no modal frequency near $3\omega_1$ or $4\omega_1$.
%

\subsection{Tests with proposed harmonization}
Tests with harmonization are restricted to the frequency range around the peak of the second response harmonic encountered in the preliminary test (\fref{prelim_test}b).
A finer frequency spacing was specified, and relatively long hold phases of \SI{5}{\second} were used, which corresponds to about one thousand excitation periods.
In accordance with the observations from the preliminary test, the truncation order was set to $H=4$, and the complete set $\mathcal H=\lbrace 2,3,4\rbrace$ was considered in the harmonization module.
A relatively low cutoff frequency of the adaptive filter was used, $\omLP = \SI{5}{\per\second}= \num{3.5e-3} \omega_1$.
The gains were tuned as proposed in \sref{gainTuning}, which led to $k_{\mrm p} = \SI{0.2}{\volt\square\second\per\meter}$ and $k_{\mrm i} = \SI{0.1}{\volt\second\per\meter}$.
Recall that those gains are kept fixed throughout the excitation frequency steps, and the gains are identical for all harmonics.
With those gains, there is no clear separation of the time scales of the fundamental and the higher harmonic controller, as opposed to the virtual experiment.
The results are shown in \fref{result_HHC}.
Besides harmonization results, test results without harmonization ($\mathcal H=\lbrace\rbrace$) are also shown.
Three consecutive test runs were done in both cases; the mean value is shown as curve, the spread as shaded area.
%
\begin{figure}[p!]
\centering
\includegraphics[]{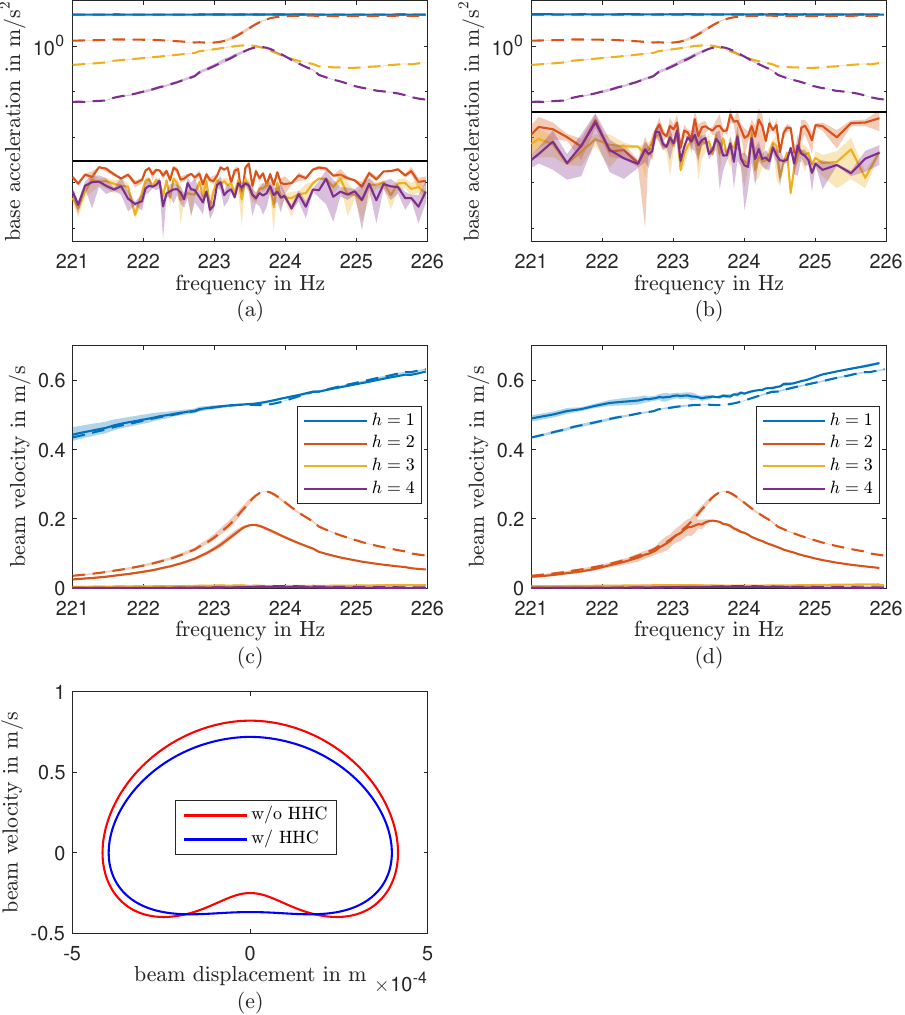}
\caption{Real experiment: Stepped sine test with (solid lines) vs. without harmonization (dashed lines); (top) base acceleration, (middle) response velocity, and (bottom) phase projection at peak of second response harmonic. Results depicted as solid lines in (a) and (c) were obtained by proposed harmonization approach, whereas those in (b) and (d) were obtained by iterative state-of-the-art approach. 
}
\label{fig:result_HHC}
\end{figure}
\\
In contrast to the preliminary tests (\fref{prelim_test}a), the fundamental excitation level remains perfectly constant (\fref{result_HHC}a).
With the proposed harmonization, the higher excitation harmonics are brought down to the noise floor (\fref{result_HHC}a).
More specifically, the magnitude of the non-fundamental Fourier coefficients remained $<0.06\%$ of the fundamental one.
The effect of the harmonization on the response can be viewed as only moderate.
Yet, it clearly exceeds the repeatability spread.
Without harmonization, the third response harmonic is larger, which is in contrast to the results of the virtual experiment shown in \fref{virt_exp_main}d.
An effect on the phase projection is also visible in \fref{result_HHC}e.

\subsection{Comparison to iterative state-of-the-art harmonization}
As state-of-the-art reference, the particular implementation of the iterative technique described in \cite{Tatzko.2023} was used.
As described in the introduction, the idea is to consider the test objective $\tilde A_{\mrm{b},1}-\hat A_{\mrm{b},1}=0$, $\tilde A_{\mrm{b},2}=0$, $\tilde A_{\mrm{b},3}=0$, $\tilde A_{\mrm{b},4}=0$, as equation system, where the Fourier coefficients of the voltage $U_1$, $U_2$, $U_3$, $U_4$ are the unknowns.
The equation system is solved using Newton-type iterations.
In the first and every second iteration, the Jacobian is approximated using finite differences, while rank-one updates as in Broyden's method are applied in the other iterations.
The latter leads to faster iterations, but leads to a less accurate estimation of the Jacobian, so that more iterations are generally required.
The iterations are stopped when the residual of the equation system is sufficiently small.
More specifically, the termination criterion was that the magnitude of the real and the imaginary parts of $\tilde A_{\mrm{b},2}$, $\tilde A_{\mrm{b},3}$ and $\tilde A_{\mrm{b},4}$ all fall below the threshold of $0.5\%~\hat A_{\mrm{b},1}$.
Between 1 and 9 iterations were needed throughout the tests carried out in the present work, with an average of about 3.
\\
Two runs of the stepped sine test with iterative harmonization were done; the mean value is shown as curve, the spread as shaded area in \fref{result_HHC}b and d.
The results obtained without harmonization (dashed lines) in \fref{result_HHC}b and d are the same as those depicted in \fref{result_HHC}a and c.
Good qualitative agreement with the results of the proposed harmonization is observed.
Some quantitative deviations appear, and a larger repeatability spread is observed for the iterative method.
This is attributed to a slow time-variability of the structure under test, related to its high thermal sensitivity.
In particular, the material temperature of the structure under test increased during longer testing, due to the heat transferred from the exciter, and the heat produced within the beam due to high-level vibrations.
Because of the clamping at both ends, the free thermal expansion is constrained, so that thermal strains induce a change of the membrane prestress, which affects both the linear and the nonlinear vibration behavior.
The wider spread observed for the iterative method, as well as the deviation of the mean are explained by the longer test duration, as detailed below.
\\
A lower residual harmonic distortion was achieved with the proposed method (\cf \fref{result_HHC}a vs. b).
It might be possible to reach a lower distortion with the iterative method by setting a tighter tolerance, but this was not checked.
In any case, this would lead to more iterations and thus a longer test duration.
The test duration was already longer for the given settings:
The iterative method took in average \SI{37}{\second} per frequency point, in contrast to \SI{6}{\second} with the proposed method. 
It should be remarked that there is potential for speedup with both methods.
In the case of the proposed method, it is expected that a faster control can be achieved by increasing the cutoff frequency, without exceeding the harmonic distortion tolerance specified for the iterative method.
In the case of the iterative method, it is expected that the hold phases of currently a few houndred excitation periods can be further reduced.
\\
The test duration of the iterative method increases substantially with the truncation order $H$.
This is due to the effort required for the finite difference approximation of the Jacobian.
The idea behind replacing this by rank-one updates in every other iteration is to alleviate this to some extent.
However, more iterations are to be expected if a less accurate approximation of the Jacobian is used.
The test duration of the proposed method, in contrast, is not expected to increase with the truncation order $H$.

\section{Conclusions\label{sec:conclusions}}
An iteration-free method for suppressing higher excitation harmonics has been proposed.
It relies on a feedback control loop that adjusts the Fourier coefficients of the voltage input to the exciter, using the respective Fourier coefficients of the applied excitation as control error.
Analytical relations between proportional and integral control gain, and parameters of the exciter-structure-system were derived for a simplified linear model.
In particular, this is useful for selecting an appropriate excitation configuration, including the drive point in the case of force excitation.
It was reasoned that a nonlinear model would be needed for an appropriate model-based control design (since the generation of higher and interaction among harmonics are essentially nonlinear phenomena), and that requiring such a model would make the method practically useless.
Thus, a heuristic tuning scheme of the control parameters was proposed.
The key parameters to be selected are the truncation order $H$, the cutoff frequency $\omLP$ of the adaptive filter that provides the steady-flow estimate of the required Fourier coefficients, and the gains of the proportional-integral controllers.
The considered examples show that these parameters can be kept fixed throughout the tested range of excitation frequencies, and the same gains can be used for all harmonics.
This is viewed as an important practical aspect.
It should be stressed that the virtual and real experiment were designed to challenge the harmonically decoupled control strategy.
In particular, this was done by driving the structure under test into the strongly nonlinear regime, and deliberately tune it into a 1:2 or 1:3 internal resonance, in order to provoke resonant interactions among the harmonically coupled modes.
Yet, excellent harmonization was achieved, where the higher harmonics were suppressed down to the noise floor, both in the force and the base excited configuration.
The controller showed robust behavior even in the presence of sudden events like a jump from one response branch to another.
The proposed method was found superior over the iterative state of the art:
It is simpler to implement, enables fast testing, independent of the truncation order, and it is easy to achieve a lower harmonic distortion.
\\
The proposed harmonization module should, in principle, be compatible with any state-of-the-art method for fundamental harmonic control.
The examples in the present work were restricted to the case of an imposed excitation frequency; the combination with (fundamental harmonic) phase and/or response control should be analyzed in the future.
Having complete control over the applied excitation, in terms of magnitude and phase of the fundamental harmonic, and the purely sinusoidal waveform, one has effectively mastered the important challenge of inevitable exciter-structure interaction.
It seems useful to implement an automatic steady state detection, because it is difficult to estimate the settling time before the test, and the settling time was found to vary to some extent.
Besides stepped sine testing, sufficiently slow frequency sweeps appear feasible, too, in the light of the high effectiveness, robustness and speed of the proposed method, which are impossible with an iterative method.
The proposed method could also be useful to achieve a multi-harmonic excitation (with specific, non-zero higher harmonics).

\section*{Acknowledgements}
This work is based on a project carried out during the \textit{Tribomechadynamics Research Camp} (TRC) 2022.
For more information, visit \url{http://tmd.rice.edu}.
The authors gratefully acknowledge the sponsoring of the TRC 2022 by MTU Aero Engines AG.
The authors acknowledge the mentorship of Ben Pacini, Sandia National Laboratories, and Ludovic Renson, Imperial College, as well as the participation of Tong Zhou, Université de Liège, during the TRC.
\par
M. Krack is grateful for the funding received by the Deutsche Forschungsgemeinschaft (DFG, German Research Foundation) [Project 402813361].
\par
The authors would like to thank dSpace GmbH for providing the MicroLabBox used to conduct the experiments.

\end{document}